\documentstyle[aps]{revtex}

\begin{document}
\title{Some comments on narrow resonances $D_{s_{1}}^{*}\left( 2.46\text{ GeV}%
/c^{2}\right) $ and $D_{s_{0}}\left( 2.317\text{ GeV}/c^{2}\right) $ }
\author{Fayyazuddin and Riazuddin}
\address{National Centre for Physics, Quaid-i-Azam University, Islamabad, Pakistan.}
\maketitle

\begin{abstract}
The newly observed resonances $D_{s_{1}}^{*}$ and $D_{s_{0}}$ are discussed
in a potential model. The relationship between the mass difference between $%
p $-states $D_{s_{1}}^{*}(1^{+})$, $D_{s_{0}}(0^{+})$ and $s$-states $%
D_{s}^{*}(1^{-})$, $D_{s}(0^{-})$ is also examined. Some remarks about the
state $D_{_{1}}^{*}(1^{+})$ and $D_{_{0}}(0^{+})$ in the non- strange
sectors are also made.
\end{abstract}

Some time back BaBar collaboration \cite{1} reported a narrow resonance with
a mass near $2.32$ GeV$/c^{2}$ which decay to $D_{s}\pi ^{0}$. Recently CLEO
collaboration \cite{2} have confirmed the state at $2.317$ GeV$/c^{2}$ and
have also reported a new resonance at $2.46$ GeV$/c^{2}$ which decays to $%
D_{s}^{*}\pi ^{0}$. Since these states decay to $D_{s}\pi ^{0}$ and $%
D_{s}^{*}\pi ^{0}$, it is natural to assign them the quantum numbers $%
J^{P}=0^{+}$ and $1^{+}$ respectively as required by angular momentum and
parity conservation. In this note we comment on the existence of relatively
light $j=1/2$ states in a potential model considered by us in $1993$\cite{3}.

In the heavy quark limit [HQET], the heavy quark spin is decoupled; it is
natural to combine $\vec{j}=\vec{L}+\vec{S}_{q}$ with $\vec{S}_{Q}$ i.e. $%
\vec{J}=\vec{L}+\vec{S}_{Q}$, where $q=u$, $d$, $s$ and $Q=c$ or $b$ \cite{4}%
. Thus for the $p$-states, we get two multiplets, one with $j=3/2$ and other
with $j=1/2$. Hence we have the multiplets for the bound states $Q\bar{q}$: $%
l=0$ [$D^{*}(1^{-})$, $D(0^{-})$], $l=1$ [$D_{2}^{*}(2^{+})$, $D_{1}(1^{+})$]%
$_{j=3/2}$, and [$D_{_{1}}^{*}(1^{+})$, $D_{0}(0^{+})$]$_{j=1/2}$. The
degeneracy between $j=3/2$ and $j=1/2$ multiplets is removed by the
spin-orbit coupling $\vec{L}\cdot \vec{S}_{q}$. The hyperfine mass splitting
between two members of each multiplet arises from the Fermi term $\vec{S}%
_{q}\cdot \vec{S}_{Q}$, the spin-orbit coupling term $\left( \vec{S}_{Q}+%
\vec{S}_{q}\right) \cdot \vec{L}$, and the tensor term \cite{5} 
\[
S_{12}\equiv \left[ \frac{12}{r^{2}}\left( \vec{S}_{q}\cdot \vec{r}\right)
\left( \vec{S}_{Q}\cdot \vec{r}\right) -4\vec{S}_{q}\cdot \vec{S}_{Q}\right] 
\]
These terms vanish in the heavy quark limit $\left( m_{Q}\rightarrow \infty
\right) $. Based on these considerations the effective Hamiltonian for the $q%
\bar{Q}$ or $Q\bar{q}$ states can be written \cite{3}: 
\begin{eqnarray}
H &=&H_{0}+\frac{1}{2m_{q}^{2}}\vec{S}_{q}\cdot \vec{L}\left[ \frac{1}{r}%
\frac{dV_{1}}{dr}\right] +\frac{1}{8m_{q}^{2}}\nabla ^{2}V_{1}+\frac{2}{%
3m_{q}m_{Q}}\vec{S}_{q}\cdot \vec{S}_{Q}\nabla ^{2}V_{2}  \nonumber
\label{01} \\
&&+\frac{1}{m_{q}m_{Q}}\left( \vec{S}_{q}+\vec{S}_{Q}\right) \cdot \vec{L}%
\left[ \frac{1}{r}\frac{dV_{2}}{dr}\right] +\frac{1}{12m_{q}m_{Q}}%
S_{12}\left[ \frac{1}{r}\frac{dV_{2}}{dr}-\frac{d^{2}V_{2}}{dr^{2}}\right]
\label{01}
\end{eqnarray}
where 
\[
H_{0}=\frac{p^{2}}{2\mu }+V\left( r\right) \text{, }\mu =\frac{m_{q}m_{Q}}{%
m_{q}+m_{Q}},\text{ }\vec{p}=-i\vec{\nabla} 
\]
The third term in eq. (\ref{01}) is the\ Darwin term. We take $V\left(
r\right) $ to be Cornell potential \cite{6} 
\begin{equation}
V\left( r\right) =\frac{r}{b^{2}}-\frac{K}{r}-A\text{ }\left[ b=2.34\text{
GeV}^{-1}\text{, }K=0.52\right]  \label{02}
\end{equation}
This potential is taken to be flavor independent. The constituent quark
masses are taken as $m_{u}=m_{d}=0.34$ GeV, $m_{s}=0.48$ GeV, $m_{c}=1.52$
GeV. We here confine ourself to the charmed quark only. The potentials $%
V_{1} $ and $V_{2}$ which occur in the spin-orbit, the Darwin, the Fermi and
the tensor terms, responsible for the fine structure are taken to be
one-gluon induced Coulomb like potential. 
\begin{equation}
V_{1}\left( r\right) =V_{2}\left( r\right) =-\frac{K^{\prime }}{r}\text{, }%
K^{\prime }=0.60  \label{3a}
\end{equation}
Based on the potential model outlined above, we discussed the mass spectrum
of $l=1$, charmed resonances in ref.\cite{3}. In particular, we predicted
two $j=1/2$ $p$-states $D_{1}^{*}(1^{+})$ and $D_{0}(0^{+})$ at $2.29$ GeV
and $2.19$ GeV respectively. These resonances are only $280$ MeV and $321$
MeV higher then the corresponding $s$-states $D^{*}(1^{-})$ and $D(0^{-})$.
But they are above the threshold for their main decay channels $D^{*}\pi $
and $D\pi $ respectively. Since these decays are $s$-wave decays in HQET;
the decay width will be large; $D_{1}^{*}$ and $D_{0}$ would appear as broad
resonances. These states have now been observed by Belle Collaboration at
some what higher mass [see below]. Thus in view of the fact that ($%
m_{s}-m_{d}=0.14$), one would expect the $D_{s_{1}}^{*}$ and $D_{s_{0}}$
states at $2.43$ GeV and $2.33$ GeV respectively. The work of ref.\cite{3}
was extended to strange and bottom quarks in ref.\cite{7}. In reference\cite
{7}, we predicted the masses of $D_{s_{1}}^{*+}$ and $D_{s_{0}}^{+}$ at $%
2.453$ GeV and $2.357$ GeV respectively about $340$ MeV and $389$ MeV higher
than the corresponding states $D_{s}^{*+}\left( 1^{-}\right) $ and $%
D_{s}^{+}\left( 0^{-}\right) $. The former agrees with the resonances $%
D_{s_{1}}^{*}(2.46$ GeV$)$in\cite{2} while the later is at somewhat higher
mass than that observed in \cite{1,2}. Since $D_{s_{1}}^{*+}$ and $%
D_{s_{0}}^{+}$ are below the threshold of the decay channels $D^{*}K$ and $%
DK $, they will appear as narrow resonances.

We now comment on the mass difference between members of $j=3/2$ and $1/2$
multiplets. In the potential model\cite{7}, we find the mass difference $%
m_{D_{s_{1}}^{*}}-m_{D_{s_{0}}}\sim 100$ MeV, whereas the experimental value
for this mass difference\cite{2} is $144$ MeV close to the mass difference $%
m_{D_{s}^{*}}-m_{D_{s}}=143.8\pm 0.4$ MeV. In order to see whether this
deficiency is an artifact of the potential model obtained in the previous
paragraph but not of the bound state picture as such we want to discuss some
general features of the model. From Eqs. (\ref{01}) and (\ref{3a}) 
\begin{eqnarray}
m_{D_{s}^{*}}-m_{D_{s}} &=&\left\langle V_{F}\right\rangle _{\text{triplet}%
}-\left\langle V_{F}\right\rangle _{\text{singlet}}  \nonumber \\
&=&\frac{2}{3m_{c}m_{s}}4\pi K^{\prime }\left| \Psi _{1s}\left( 0\right)
\right| ^{2}\equiv \frac{2}{3}\lambda _{s}  \label{04}
\end{eqnarray}
\begin{eqnarray}
m_{D_{s_{1}}^{*}}-m_{D_{s_{0}}} &=&\frac{1}{m_{c}m_{s}}\left[ \left( \vec{S}%
\cdot \vec{L}\right) _{D_{s_{1}}^{*}}-\left( \vec{S}\cdot \vec{L}\right)
_{D_{s_{0}}}\right] K^{\prime }\left\langle \frac{1}{r^{3}}\right\rangle
_{1p}+\frac{1}{12m_{c}m_{s}}\left[ \left( S_{12}\right)
_{_{D_{s_{1}}^{*}}}-\left( S_{12}\right) _{D_{s_{0}}}\right] 3K^{\prime
}\left\langle \frac{1}{r^{3}}\right\rangle _{1p}  \nonumber \\
&=&\frac{8}{3m_{c}m_{s}}K^{\prime }\left\langle \frac{1}{r^{3}}\right\rangle
_{1p}\equiv \frac{8}{3}\lambda _{1s}^{\prime }  \label{05}
\end{eqnarray}
\begin{eqnarray}
m_{D_{s_{2}}^{*}}-m_{D_{s_{1}}} &=&\frac{1}{m_{c}m_{s}}\left[ \left( \vec{S}%
\cdot \vec{L}\right) _{D_{s_{2}}^{*}}-\left( \vec{S}\cdot \vec{L}\right)
_{D_{s_{1}}}\right] K^{\prime }\left\langle \frac{1}{r^{3}}\right\rangle
_{1p}+\frac{1}{12m_{c}m_{s}}\left[ \left( S_{12}\right)
_{D_{s_{2}}^{*}}-\left( S_{12}\right) _{D_{s_{1}}}\right] 3K^{\prime
}\left\langle \frac{1}{r^{3}}\right\rangle _{1p}  \nonumber \\
&=&\frac{16}{15}\frac{1}{m_{c}m_{s}}K^{\prime }\left\langle \frac{1}{r^{3}}%
\right\rangle _{1p}\equiv \frac{16}{15}\lambda _{1s}^{\prime }  \label{05a}
\end{eqnarray}
\begin{equation}
m_{D_{s_{1}}^{*}}-m_{D_{s_{0}}}=\frac{5}{2}(m_{D_{s2}^{*}}-m_{D_{s1}})
\label{05b}
\end{equation}
As is well known Eq. (\ref{04}) is on a solid ground. The mass splitting
between $^{3}S_{1}$ and $^{1}S_{0}$ state is due to Fermi interaction $\bar{%
\mu}_{q}\cdot \bar{\mu}_{Q}$ and is goverened by the short range Coulomb
like potential. For $p$- wave states, the same interaction induces the
tensor term $S_{12}$. The spin-orbit coupling $\vec{S}\cdot \vec{L}$ is
needed to preserve the observed hierarchy in the mass spectrum of heavy $p$%
-state mesons. In its absence, we would get $%
m_{D_{s_{_{1}}}^{*}}-m_{D_{s_{_{0}}}}=-5(m_{D_{s_{_{2}}}^{*}}-m_{D_{s_{_{1}}}}) 
$ in contradiction to experimentally observed mass heirarchy. The spin orbit
coupling $\frac{1}{m_{q}^{2}}\vec{S}_{q}\cdot \vec{L}$ gives the mass
splitting 
\begin{equation}
m_{j=3/2}-m_{j=1/2}=\frac{3}{2}\frac{1}{2m_{q}^{2}}\left\langle \frac{1}{r}%
\frac{dV_{1}}{dr}\right\rangle =\frac{3}{4}\frac{m_{c}}{m_{q}}\lambda _{1q}
\label{05c}
\end{equation}
With confining potential $V(r)$ given in Eq. (\ref{02}), one has the
relations \cite{8}

\begin{equation}
4\pi \left| \Psi _{1s}\left( 0\right) \right| ^{2}=2\mu \left\langle \frac{dV%
}{dr}\right\rangle _{1s}=2\mu \left[ \frac{1}{b^{2}}+K\left\langle \frac{1}{%
r^{2}}\right\rangle _{1s}\right]  \label{06}
\end{equation}
\begin{equation}
\left\langle \frac{1}{r^{3}}\right\rangle _{1p}=\frac{2\mu }{4}\left\langle 
\frac{dV}{dr}\right\rangle _{1p}=\frac{2\mu }{4}\left[ \frac{1}{b^{2}}%
+K\left\langle \frac{1}{r^{2}}\right\rangle _{1p}\right]  \label{07}
\end{equation}
Using these relations we get from Eqs. (\ref{04}) and (\ref{05}) 
\begin{eqnarray}
m_{D_{s}^{*}}-m_{D_{s}} &=&\frac{2}{3}K^{\prime }\frac{2\mu }{m_{c}m_{s}}%
\left[ \frac{1}{b^{2}}+K\left\langle \frac{1}{r^{2}}\right\rangle
_{1s}\right]  \label{08} \\
m_{D_{s_{1}}^{*}}-m_{D_{s_{0}}} &=&\frac{2}{3}K^{\prime }\frac{2\mu }{%
m_{c}m_{s}}\left[ \frac{1}{b^{2}}+K\left\langle \frac{1}{r^{2}}\right\rangle
_{1p}\right]  \label{09}
\end{eqnarray}
Hence we obtain 
\begin{equation}
m_{D_{s_{1}}^{*}}-m_{D_{s_{0}}}=m_{D_{s}^{*}}-m_{D_{s}}  \label{11}
\end{equation}
only if 
\begin{equation}
\left\langle \frac{1}{r^{2}}\right\rangle _{1p}=\left\langle \frac{1}{r^{2}}%
\right\rangle _{1s}  \label{13}
\end{equation}
It is unlikely that this equality would hold in a potential model. In fact
one would expect $\left\langle \frac{1}{r^{2}}\right\rangle _{1p}$ to be
less than $\left\langle \frac{1}{r^{2}}\right\rangle _{1s}$. This is the
reason why we get $\left( m_{D_{s_{1}}^{*}}-m_{D_{s_{0}}}\right) $ less than 
$\left( m_{D_{s}^{*}}-m_{D_{s}}\right) $. In refrence \cite{3,7}, we
obtained 
\begin{eqnarray}
\left\langle \frac{1}{r^{2}}\right\rangle _{1s} &\approx &0.357\text{ GeV}%
^{2}  \nonumber \\
\left\langle \frac{1}{r^{2}}\right\rangle _{1p} &\approx &0.088\text{ GeV}%
^{2}  \label{14}
\end{eqnarray}
These values give 
\begin{eqnarray}
m_{D_{s}^{*}}-m_{D_{s}} &\approx &147\text{ MeV}  \nonumber \\
m_{D_{s_{1}}^{*}}-m_{D_{s_{0}}} &\approx &91\text{ MeV}  \label{15}
\end{eqnarray}
We now wish to comment on recently observed states $D_{1}^{*}$ and $D_{0}$
in the non-strange sector by the Belle Collabration\cite{Belle} 
\begin{eqnarray}
m_{D_{1}^{*}} &=&(2427\pm 26\pm 20\pm 15)\text{ MeV/}c^{2}  \nonumber \\
m_{D_{0}} &=&(2308\pm 17\pm 15\pm 28)\text{ MeV/}c^{2}  \label{15a}
\end{eqnarray}
The axial vector states mix. The mixing angle is given by\cite{Belle} 
\begin{equation}
\omega =-0.10\pm 0.03\pm 0.02\pm 0.02  \label{15b}
\end{equation}
Taking these masses on their face values 
\begin{equation}
m_{D_{1}^{*}}-m_{D_{0}}\approx 119\text{ MeV; }m_{D_{1}}-m_{D_{1}^{*}}%
\approx 0  \label{15c}
\end{equation}
First we wish to show the observed mass spectra of $D$ mesons imply that the
parameters $\lambda _{q}$, $\lambda _{1q}^{\prime }$ are independent of
light flavor. Thus 
\begin{equation}
m_{D_{s}^{*}}-m_{D_{s}}\approx 144\text{ MeV; }m_{D^{*}}-m_{D}\approx 140%
\text{ MeV}\Rightarrow \lambda _{s}\approx \lambda _{d}  \label{15d}
\end{equation}
\begin{equation}
m_{D_{s_{2}}^{*}}-m_{D_{s_{_{1}}}}\approx 40\text{ MeV; }%
m_{D_{2}^{*}}-m_{D_{1}}\approx 40\text{ MeV}\Rightarrow \frac{16}{15}\lambda
_{1s}^{\prime }\approx \frac{16}{15}\lambda _{1d}^{\prime }  \label{15e}
\end{equation}
\begin{equation}
\lambda _{1s}^{\prime }=\lambda _{1d}^{\prime }=36\text{ MeV}  \label{15f}
\end{equation}
In view of the defination of $\lambda _{1q}$ given in Eq. (\ref{05c}), one
would expect $\lambda _{1q}$ to be independent of light flavor (see below).
We note that 
\begin{equation}
m_{D_{s_{_{1}}}}-m_{D_{s_{_{1}}}^{*}}=\frac{3}{4}\frac{m_{c}}{m_{s}}\lambda
_{1s}+\frac{1}{6}\lambda _{1s}^{\prime }  \label{15g}
\end{equation}
The spin orbit potential $V_{1}\left( r\right) $, for light quark in the
HQET as given in Eq. (\ref{01}) is expected to be of same form as $%
V_{2}\left( r\right) $. We may further assume $V_{1}\left( r\right) $ to
have the same strength as $V_{2}\left( r\right) $\cite{3,7}. With this
assumption 
\begin{equation}
\lambda _{1q}=\lambda _{1q}^{\prime }\approx 36\text{ MeV}  \label{15h}
\end{equation}
Thus we get 
\begin{eqnarray}
m_{D_{s_{_{1}}}}-m_{D_{s_{_{1}}}^{*}} &\approx &91\text{ MeV\thinspace
\thinspace \thinspace \thinspace \thinspace \thinspace \thinspace \thinspace
\thinspace (}75\text{ Mev\cite{2})}  \nonumber \\
m_{D_{1}}-m_{D_{1}^{*}} &\approx &126\text{ MeV}  \nonumber \\
m_{D_{1}^{*}} &=&m_{D_{1}}-126\text{ MeV}=2.295\text{ GeV}  \nonumber \\
m_{D_{0}} &\approx &2.195\text{ GeV}  \label{15i}
\end{eqnarray}
compatible with those of ref.\cite{7}. However, if we take $\lambda _{1q}$
slightly less than $\lambda _{1q}$; say $30$ MeV i.e. $V_{1}\left( r\right) $
is slightly weaker than $V_{2}\left( r\right) $: then we get 
\begin{eqnarray}
m_{D_{s_{_{1}}}}-m_{D_{s_{_{1}}}^{*}} &\approx &77\text{ MeV}  \nonumber \\
m_{D_{1}}-m_{D_{1}^{*}} &\approx &106\text{ MeV}  \nonumber \\
m_{D_{1}^{*}} &=&2.315\text{ GeV}  \nonumber \\
m_{D_{0}} &=&2.215\text{ GeV}  \label{15j}
\end{eqnarray}
Finally, the mixing angle between axial vector states $D_{1}$ and $D_{1}^{*}$
is given by 
\begin{eqnarray}
\tan 2\omega &=&\frac{2m_{D_{1}-D_{1}^{*}}}{m_{D_{1}}-m_{D_{1}^{*}}} 
\nonumber \\
&=&-\frac{\frac{\sqrt{2}}{3}\lambda _{1d}^{\prime }}{\frac{3}{4}\frac{m_{c}}{%
m_{s}}\lambda _{1d}+\frac{1}{6}\lambda _{1d}^{\prime }}  \nonumber \\
&\approx &0.135  \label{15k}
\end{eqnarray}
\begin{equation}
\omega =0.07  \label{15l}
\end{equation}
to be compared with the value of $\omega $ \cite{Belle} given in Eq. (\ref
{15b}). Higher value of $\omega $ would imply $\lambda _{1d}$ less than $%
\lambda _{1d}^{\prime }$ viz $V_{1}\left( r\right) $ is weaker than $%
V_{2}\left( r\right) $. For $\lambda _{1d}\approx 30$ MeV, we get $\omega
=0.08$.

To reinforce the above conclusion, we note from Eqs. (\ref{05}) (\ref{05a})
and (\ref{08}) 
\begin{equation}
\frac{5m_{D_{s_{2}}^{*}}+3m_{D_{_{s_{1}}}}}{8}-\frac{%
3m_{D_{s_{_{1}}}^{*}}+m_{D_{s_{_{0}}}}}{4}=\frac{3}{2}\left( \frac{m_{c}}{%
m_{s}}\lambda _{1s}+\lambda _{1s}^{\prime }\right)  \label{15m}
\end{equation}
Using the experimental values for masses \cite{1,2}, we get 
\begin{equation}
\frac{3}{2}\left( \frac{m_{c}}{m_{s}}\lambda _{1s}+\lambda _{1s}^{\prime
}\right) =133\text{ MeV}  \label{15n}
\end{equation}
Now using $\lambda _{1s}^{\prime }\approx 36$ MeV, we get, $\lambda
_{1s}\approx 33$ MeV compatible with the assumption stated above. Flavor
independence of these parameters give 
\begin{eqnarray}
\frac{3m_{D_{1}^{*}}+m_{D_{0}}}{4} &=&\frac{5m_{D_{2}^{*}}+3m_{D_{1}}}{5}-133%
\text{ MeV}  \nonumber \\
&\approx &2312\text{ MeV}  \label{15o}
\end{eqnarray}
This value is about $85$ MeV, below that of Belle Collaboration, but only
about $30$ MeV above the values implied by Eq. (\ref{15i}). However this
value is in agreement with that obtained from Eq. (\ref{15j}).

The following comments are in order. In reference \cite{9,10}, it was
suggested that in the heavy quark limit $j=1/2$ states with $J^P=1^{+}$ and $%
0^{+}$ are chiral partners of $1^{-}$ and $0^{-}$. In ref. \cite{9} mass
difference between parity doublets $\left( 0^{-}\text{,}1^{-}\right) $ and $%
\left( 0^{+}\text{,}1^{+}\right) $ arises due to chiral symmetry breaking
and thus expected to be small, a feature which we also get. In particular
they find the mass difference of order $338$ MeV. In ref. \cite{10}, they
obtained $m_{D_1^{*}}-m_{D_0}\approx m_{D^{*}}-m_D$. In the bound state
model, we get $m_{D_0}-m_D\approx 326$ MeV. Further we note that $j=1/2$ $p$%
-wave multiplet lie below $j=3/2$ multiplet, due to spin-orbit coupling of
light quark in the limit of heavy quark spin symmetry reminiscencet of fine
structure of hydrogen atom spectrum.

We now briefly discuss decays of resonances $D_{s1}^{*}$ and $D_{s0}$.The
isospin conserving decays 
\begin{eqnarray*}
D_{s_{1}}^{*} &\longrightarrow &D^{*0}K^{+}(D^{*+}K^{0}) \\
D_{s_{0}}^{+} &\longrightarrow &D^{0}K^{+}(D^{+}K^{0})
\end{eqnarray*}
are not energetically allowed. The experimentally observed decays $%
D_{s}^{*+}\pi ^{0}$ and $D_{s}^{+}\pi ^{0}$ violate isospin. One obvious
possibility is that these decays proceed via $\eta $-meson : 
\begin{eqnarray*}
D_{s_{1}}^{*} &\longrightarrow &D_{s}^{*+}\eta \longrightarrow D_{s}^{*+}\pi
^{0} \\
D_{s_{0}}^{+} &\longrightarrow &D_{s}^{+}\eta \longrightarrow D_{s}^{+}\pi
^{0}
\end{eqnarray*}

In this picture the coupling $g_{D_{s0}D_{s}\pi }$ can be expressed in terms
of $g_{D_{s0}D_{s}\eta }$ as: 
\begin{equation}
g_{D_{s_{0}}D_{s}\pi }=g_{D_{s_{0}}D_{s}\eta }\frac{m_{\eta -\pi ^{0}}^{2}}{%
m_{\eta }^{2}-m_{\pi ^{0}}^{2}}  \label{16}
\end{equation}
where\cite{11}

\begin{equation}
m_{\eta -\pi ^0}^2=-\frac 1{\sqrt{3}}\left[ \left(
m_{K^0}^2-m_{K^{+}}^2\right) +\left( m_{\pi ^{+}}^2-m_{\pi ^0}^2\right)
\right]  \label{17}
\end{equation}
Then using $SU(3)$

\begin{equation}
g_{D_{s_{0}}^{+}D_{s}^{+}\eta }=-\sqrt{\frac{2}{3}}g_{D_{0}^{+}D^{+}\pi ^{0}}
\label{18}
\end{equation}
we get

\begin{equation}
\left( \frac{g_{D_{s_{0}}^{+}D_{s}^{+}\pi ^{0}}}{g_{D_{0}^{+}D^{+}\pi }}%
\right) ^{2}=\frac{2}{3}\left( \frac{m_{\eta -\pi ^{0}}^{2}}{m_{\eta
}^{2}-m_{\pi }^{2}}\right) ^{2}\approx 7.7\times 10^{-5}  \label{19}
\end{equation}

\begin{eqnarray}
\frac{\Gamma \left( D_{s_{0}}^{+}\longrightarrow D_{s}^{+}\pi ^{0}\right) }{%
\Gamma \left( D_{0}^{+}\longrightarrow D^{+}\pi ^{0}\right) } &=&\frac{%
\left| \vec{p}\right| _{D_{s}}}{\left| \vec{p}\right| _{D}}\frac{%
m_{D_{0}}^{2}}{m_{D_{s0}}^{2}}\left[ \frac{\left(
m_{D_{s_{0}}}-m_{D_{s}}\right) }{\left( m_{D_{0}}-m_{D}\right) }\right]
^{2}\times \left[ 7.7\times 10^{-5}\right]  \label{20} \\
&\approx &\left( 1.07\right) \left( 7.7\times 10^{-5}\right) =8.2\times
10^{-5}  \label{21}
\end{eqnarray}
where we have defined the dimensionless coupling constant $%
g_{D_{0}^{+}D^{+}\pi ^{0}}$ as\cite{12}

\begin{eqnarray}
M &=&\sqrt{4p_0p_0^{\prime }}\left\langle D^{+}(p^{\prime })\left| J_\pi
\right| D_0(p)\right\rangle =i\left( \frac{m_{D_0}^2-m_D^2}{2m_D}\right)
g_{D_0^{+}D^{+}\pi ^0}  \nonumber  \label{22} \\
&\approx &i\left( m_{D_0}-m_D\right) g_{D_0^{+}D^{+}\pi ^0}  \label{22}
\end{eqnarray}
and have used $m_{D_0}=2.20$ GeV and the experimental values for other
masses. Assuming the decay width, 
\begin{equation}
\Gamma \left( D_0^{+}\longrightarrow D^{+}\pi ^0\right) \sim 200\,\text{MeV}
\label{23}
\end{equation}
we get

\begin{equation}
\Gamma \left( D_{s_{0}}^{+}\longrightarrow D_{s}^{+}\pi ^{0}\right) \sim 16\,%
\text{keV}  \label{24}
\end{equation}
That the decay width $\Gamma \left( D_{0}^{+}\longrightarrow D^{+}\pi
^{0}\right) $ is of order 200 MeV can be seen as follows: In HQET, the
coupling $g_{D^{*}D\pi }$ is usually parametrized as $g_{D^{*}D\pi }=\lambda
_{D}m_{D}/f_{\pi }$; in the same spirt $g_{D_{0}D\pi }$ is parameterized as%
\cite{12}:

\begin{eqnarray}
&&\left. g_{D_{0}D\pi }=2\lambda _{D_{0}}\frac{m_{D}}{f_{\pi }}\right.
\label{25} \\
&&\left[ f_{\pi }=132\,\text{MeV}\right]  \nonumber
\end{eqnarray}
Thus we get

\begin{eqnarray}
\Gamma \left( D_0^{+}\longrightarrow D^{+}\pi ^0\right) &=&\frac 1{8\pi }%
\left| M\right| ^2\left| \vec{p}\right| \frac 1{m_{D_0}^2}=\frac 1{8\pi }%
\frac 1{m_{D_0}^2}\frac{\left( 4\lambda _{D^0}^2m_D^2\right) }{f_\pi ^2}%
\left( m_{D_0}-m_D\right) ^2\left| \vec{p}\right|  \nonumber \\
&\approx &202\lambda _{D^0}^2\text{ MeV}\leq 202\text{ MeV}  \label{26}
\end{eqnarray}

In HQET the decay 
\[
D_{s_{_{1}}}^{*+}\rightarrow D_{s}^{*+}\pi ^{0} 
\]
is also $s$-wave and is related to 
\[
D_{s_{_{0}}}^{+}\longrightarrow D_{s}^{+}\pi ^{0} 
\]
as follows:

\begin{equation}
\frac{\Gamma \left( D_{s_{1}}^{*}\rightarrow D_{s}^{*+}\pi ^{0}\right) }{%
\Gamma \left( D_{s_{0}}^{+}\rightarrow D_{s}^{+}\pi ^{0}\right) }=\frac{%
\left| \vec{p}\right| _{D_{s}^{*}}}{\left| \vec{p}\right| _{D_{s}}}\frac{%
\left( m_{D_{s_{_{1}}}^{*}}-m_{D_{s}^{*}}\right) ^{2}}{\left(
m_{D_{s_{0}}}-m_{D_{s}}\right) ^{2}}  \label{27}
\end{equation}
Thus except for phase space, the decay width of $D_{s_{_{1}}}^{*+}%
\rightarrow D_{s}^{*+}\pi ^{0}$ is equal to that of $D_{s_{_{0}}}^{+}%
\longrightarrow D_{s}^{+}\pi ^{0}$. However, the mixing between axial vector
states $D_{s_{_{1}}}$ and $D_{s_{_{1}}}^{*}$ can contribute to the decay
width of $D_{s_{_{1}}}^{*}$. The contribution to the decay width of $%
D_{s_{1}}^{*}$ due to mixing is $\omega ^{2}\Gamma _{D_{s_{1}}}$. Since $%
\Gamma _{D_{s_{1}}}<2.3$ MeV we get $\omega ^{2}\Gamma _{D_{s_{1}}}<23$ KeV
for the mixing angle $\omega \simeq 0.1$ as implied by Eq. (\ref{15l}). Thus
width of $D_{s_{1}}^{*}$ is expected to be about twice that of $D_{s_{_{0}}}$%
.

Finally the resonances $D_{s_{1}}^{*+}$ and $D_{s_{0}}^{+}$ can also decay
to $D_{s}^{*+}\gamma $ and $D_{s}^{+}\gamma $ respectively by E1 transition%
\cite{3}.For the decay 
\[
D_{s_{_{0}}}^{+}\rightarrow D_{s}^{+}\gamma 
\]
the deacy width is given by

\begin{equation}
\Gamma =\frac{4\alpha }3\left| M^{+}\right| ^2k^3  \label{28}
\end{equation}
In the quark model\cite{3}

\begin{equation}
M^{+}=\mu \left[ \frac 2{3m_c}I_c-\frac 1{3m_s}I_s\right]  \label{29}
\end{equation}
where $I_c$ and $I_s$ are overlap integrals. Our estimate for the radiative
decay width comes out to be $0.2$ KeV\cite{7}.

To conclude, in a picture in which $D$ mesons are regarded as bound states $c%
\bar{q}$, the potential model considered by us give 
\begin{eqnarray}
m_{j=1/2} &<&m_{j=3/2}  \label{30} \\
m_{D_{s_{_{1}}}^{*}}-m_{D_{s_{0}}} &=&\frac{5}{2}\left(
m_{D_{s_{_{2}}}^{*}}-m_{D_{s_{_{1}}}}\right)  \label{31} \\
m_{D_{s_{_{1}}}^{*}}-m_{D_{s_{_{0}}}} &\approx &\frac{1}{\sqrt{2}}\left(
m_{D_{s}^{*}}-m_{D_{s}}\right)  \label{32}
\end{eqnarray}
Further, using the experimental data of refrences \cite{1,2}, our analysis
gives 
\begin{eqnarray}
\frac{3m_{D_{1}^{*}}+m_{D_{0}}}{4} &\approx &2.312\text{ GeV, }%
m_{D_{1}^{*}}=2.35\text{ GeV, }m_{D_{0}}=2.20\text{ GeV}  \nonumber \\
&&  \label{33} \\
\left( \frac{3m_{D_{1}^{*}}+m_{D_{0}}}{4}\right) -\left( \frac{%
3m_{D^{*}}+m_{D}}{4}\right) &\approx &337\text{ MeV}  \label{34}
\end{eqnarray}
The $p$-wave $j=1/2$ multiplet is $337$ MeV, above the $s$-wave multiplet.
Except for the mass relation (\ref{32}), (which is hard to understand in a
bound state picture) the general features of bound state picture are
compatible with the experimental data.

{\bf Acknowldgement}

This work was supported by a grant from the Pakistan Council of Science and
technology.


\begin{references}
\bibitem{1}  BaBar Collaboration, B.Aubert et.al. Phys.Rev.Lett.{\bf 90},
242001(2003)

\bibitem{2}  CLEO collaboration, D.Besson et.al.hep-ex/0305100

\bibitem{3}  Fayyazuddin and Riazuddin, Phys.Rev.D{\bf 48} 2224(1993)

\bibitem{4}  N.Isgur and M.B.Wise,Phys.Lett.B{\bf 232}%
,113(1989);Phys.Rev.Lett.{\bf 66},1130(1991) Ming-Lu,M.B.Wise and
N.Isgur,Phys.Rev.D{\bf 45},1553(1992)

\bibitem{5}  A. De Rujula, H. Georgi and S. Glashow, Phys. Rev. D{\bf 12}
147 (1975); E. Eichten and F.Feniberg ; Phys. Rev. D{\bf 23}, 2724 (1981),D
. Gromes ,Z. Phys. C{\bf 26}, 401(1984). J. L. Rosner,Comments Nucl. Part.
Phys. A{\bf 16}, 109 (1986)

\bibitem{6}  E.Eichten et.al,Phys.Rev.D{\bf 17},3090(1978);ibid 21,203(1980)

\bibitem{7}  Fayyazuddin and Riazuddin,J.Phys.G:Nucl.Part.{\bf 24},23(1998)

\bibitem{8}  See,for example,C.Quigg and J.L.Rosner,Phys.Rep{\bf 56}%
,167(1979)

\bibitem{9}  W.A.Bardeen and C.T.Hill,Phys.Rev.D{\bf 49} 409(1994)

\bibitem{10}  M.A.Nowak,M.Rho and I.Zahed,Phys.Rev.D{\bf 48},4370(1993)

\bibitem{Belle}  Belle Collaboration, K. Abe et al. hep-ex/0307021

\bibitem{11}  See for example: S. Weinberg, in A Festschrift for I.I. Rabi
(New York Academy of Sciences, New York, (1978))

\bibitem{12}  Fayyazuddin nad Riazuddin,Phys.Rev.D{\bf 49},3385(1994)
\end{references}
\end{document}